\documentclass[submission, Proceedings,a4paper]{SciPost}

\hypersetup{
    colorlinks,
    linkcolor={red!50!black},
    citecolor={blue!50!black},
    urlcolor={blue!80!black}
}

\usepackage[bitstream-charter]{mathdesign}
\usepackage{lineno}
\usepackage{cancel}
\DeclareSymbolFont{usualmathcal}{OMS}{cmsy}{m}{n}
\DeclareSymbolFontAlphabet{\mathcal}{usualmathcal}
\begin{document}

\begin{center}{\Large \textbf{
Feasibility of tau g-2 measurements \\ in ultra-peripheral collisions of heavy ions
}}\end{center}

\begin{center}
E. Kryshen\textsuperscript{1$\star$},
N. Burmasov\textsuperscript{1},
P. B\"uhler\textsuperscript{2} and
R. Lavicka\textsuperscript{2}
\end{center}

\begin{center}
{\bf 1} Petersburg Nuclear Physics Institute named by B. P. Konstantinov of National Research Center <<Kurchatov Institute>>, 188300, 1 mkr. Orlova roshcha, Gatchina, Russia
\\
{\bf 2} Stefan Meyer Institute for Subatomic Physics, Kegelgasse 27, 1030 Vienna,\,Austria
\\
* evgeny.kryshen@cern.ch
\end{center}

\begin{center}
\today
\end{center}


\definecolor{palegray}{gray}{0.95}
\begin{center}
\colorbox{palegray}{
  \begin{minipage}{0.95\textwidth}
    \begin{center}
    {\it  16th International Workshop on Tau Lepton Physics (TAU2021),}\\
    {\it September 27 – October 1, 2021} \\
    \doi{10.21468/SciPostPhysProc.?}\\
    \end{center}
  \end{minipage}
}
\end{center}

\section*{Abstract}
{\bf
The anomalous magnetic moment of the tau lepton, $a_\tau=(g_\tau - 2)/2$, is a sensitive probe of new physics but is extremely difficult to measure precisely in contrast to electron and muon moments. The best experimental limits were set by the DELPHI collaboration more than 15 years ago in studies of the ditau production in the $ee\to ee\tau\tau$ process. Ultra-peripheral collisions (UPCs) of heavy ions at the LHC may provide a unique opportunity to improve the $a_\tau$ constraints in the studies of ${\rm Pb + Pb}\to {\rm Pb + Pb}+\tau\tau$ process. We review recent proposals to study ditau production via semi-leptonic tau decays in Pb-Pb UPC with the available ATLAS and CMS data and discuss the feasibility to explore this process down to low transverse momenta of decay leptons with the ALICE and LHCb experiments.
}

\vspace{10pt}
\noindent\rule{\textwidth}{1pt}
\tableofcontents\thispagestyle{fancy}
\noindent\rule{\textwidth}{1pt}
\vspace{10pt}

\section{Introduction}
\label{sec:intro}

Precision measurements of the tau anomalous magnetic moment $a_\tau=(g_\tau - 2)/2$ are important for verification of QED predictions and searches for physics beyond the Standard Model (BSM), such as lepton compositeness and various supersymmetric scenarios~\cite{Martin:2001st}. Experimental  $a_\tau$ measurements are scarce since the standard spin precession methods are not applicable for $a_\tau$ measurements due to the very short $\tau$ lifetime. The best limits on $a_\tau$ were set by the DELPHI collaboration in the studies of the ditau production in the $ee\to ee\tau\tau$ process~\cite{DELPHITauLimits}:
\begin{equation}
	-0.052 < a_{\tau} < 0.013 \, (95\%\,{\rm  CL}).
\end{equation}
However there was no progress on experimental measurements of $a_\tau$ for more than 15 years.

Latest results of the Muon $g-2$ Collaboration show a significant discrepancy of four standard deviations of the measured muon anomalous magnetic moment $a_\mu$ from the Standard Model predictions revealing possible signs of BSM contributions~\cite{FermilabGm2}. The tau anomalous magnetic moment is expected to be $m_\tau^2/m_\mu^2 \sim 280$ times more sensitive to BSM physics compared to muon $g-2$, therefore precision measurements of the $\tau$ anomalous magnetic moment acquired particular interest. 

Ultra-peripheral collisions (UPCs) of heavy ions may provide a unique opportunity to improve the existing limits of the $\tau$ anomalous magnetic moment~\cite{Beresford,DYNDAL2020135682}. UPCs are characterized by impact parameters larger than the sum of radii of incoming nuclei~\cite{BaltzUpc,ContrerasUpc} therefore  electromagnetic processes, such as photon-photon interactions, become dominant in these collisions. Electromagnetic fields of incoming heavy ions can be described in terms of an equivalent photon flux whose intensity is proportional to the squared nuclear charge $Z^2$. Therefore photon-photon interactions in UPCs scale as $Z^4$ resulting in high cross sections for dilepton pair production and other processes.  

In this contribution, we review recent proposals on the tau anomalous magnetic moment measurement via ditau production studies in UPCs with ATLAS and CMS experiments and discuss the feasibility to explore this process down to low transverse momenta of the decay leptons with ALICE and LHCb detectors.

\section{Event selection strategy}

\begin{figure}[ht]
	\centering
	\includegraphics[width=0.30\textwidth]{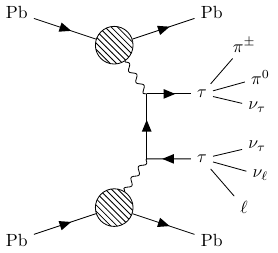}
	\caption{Production and typical decays of tau lepton pairs in Pb--Pb UPC.}
	\label{fig:ditauProduction}
\end{figure}

\begin{figure}[h]
	\centering
	\includegraphics[width=0.45\textwidth]{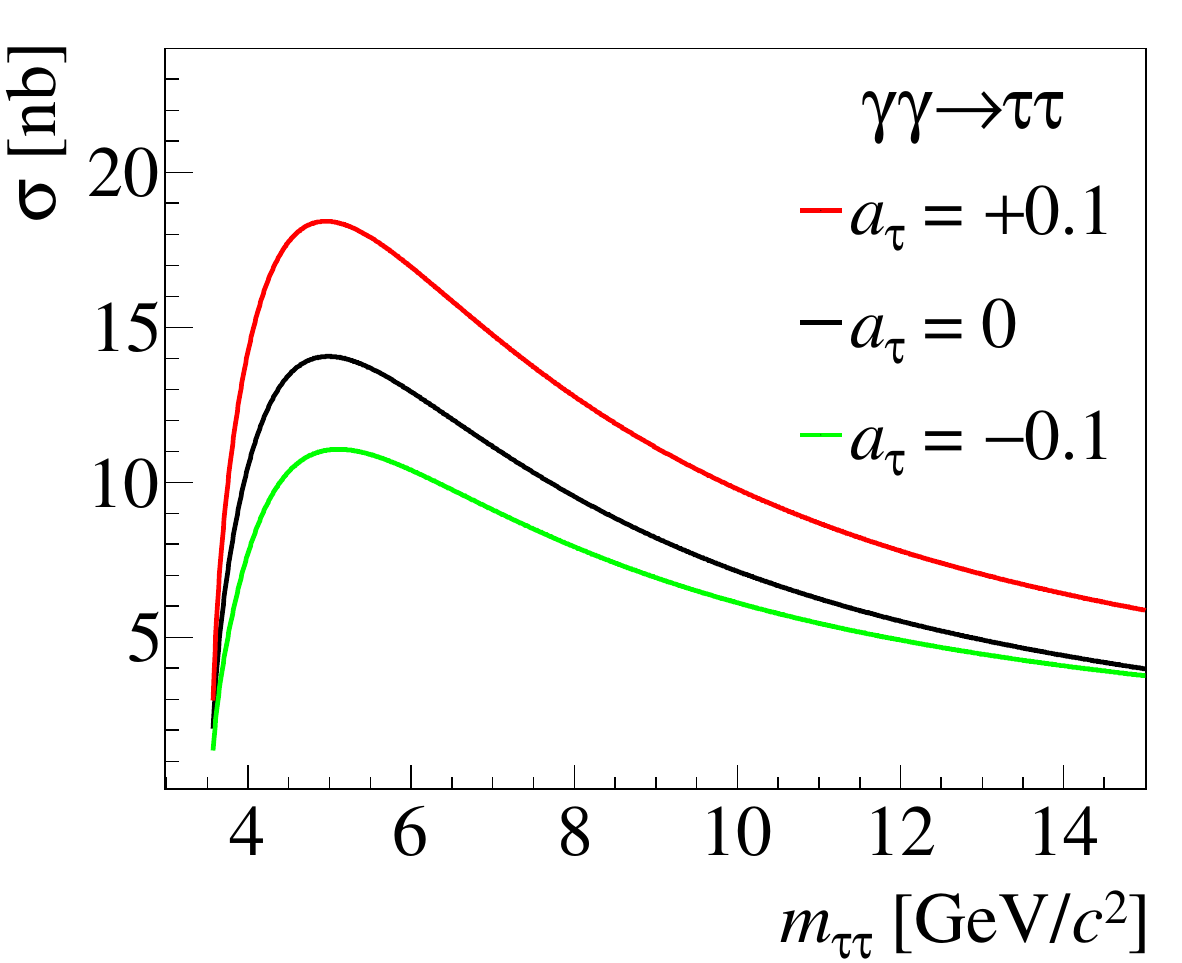}
	\caption{Elementary $\gamma\gamma\hspace{-1mm}\to\hspace{-1mm}\tau\tau$ cross sections as functions of the ditau invariant mass for different $a_{\tau}$ values. Figure from~\cite{Burmasov:2021phy}.}.
	\label{fig:elementary_cs}
\end{figure}

The process of ditau pair production in Pb--Pb UPCs is illustrated in Fig.~\ref{fig:ditauProduction}. Tau leptons are produced in the photon fusion process $\gamma\gamma \to \tau\tau$ with photon-tau vertices being sensitive to the value of the tau anomalous magnetic moment. Sensitivity of the $\gamma\gamma \to \tau\tau$ cross section to the $a_\tau$ value is illustrated in Fig.~\ref{fig:elementary_cs}.

Due to the short lifetime, taus decay into lighter leptons or hadrons before any direct interaction with the detector material. Therefore, in order to study tau pair production process in UPCs, one has to record events with a few charged decay products of tau leptons in an otherwise empty detector. In 35\% of cases a tau decays into a lighter lepton accompanied by neutrinos: $\tau^\pm \to \ell^\pm + \nu_\ell + \nu_\tau$~\cite{ParticleDataGroup:2020ssz}. Taus can also decay into one or three charged pions accompanied by neutrals with the following branching ratios: ${\rm BR}(\tau^\pm \to \pi^\pm + n\pi^0 + \nu_\tau) = 45.6\%$, ${\rm BR}(\tau^\pm \to \pi^\pm + \pi^\mp + \pi^\pm + n\pi^0 + \nu_\tau) = 19.4\%$. 

The UPC event selection strategy in LHC experiments usually relies on online trigger requirements such as the presence of one or several leptons with $p_{\rm T} \gtrsim 1$ GeV/$c$. Therefore, possible ditau selection strategy can be based on the requirement of a leading lepton from one tau decay accompanied by one or three charged tracks from the other tau.

\section{ATLAS and CMS potential}

The feasibility of $a_\tau$ measurements in the ATLAS~\cite{ATLAS:2008xda} and CMS~\cite{CMS:2008xjf} experiments were studied by two groups~\cite{Beresford,DYNDAL2020135682}. L. Beresford and J. Liu~\cite{Beresford} performed calculations within the standard model effective field theory (SMEFT), while M. Dynda\l~et al.~\cite{DYNDAL2020135682} used a more conservative approach based on the generalized photon-lepton vertex function. Both groups considered a possibility to measure $p_{\rm T}$ distributions of leading leptons with high $p_{\rm T}$ thresholds ($p_{\rm T}\gtrsim4$~GeV/$c$), accompanied by one or three charged tracks, and provided projections on $a_\tau$ limits which can be achieved by ATLAS and CMS in Pb--Pb UPCs at $\sqrt{s_{\rm NN}} = 5.02$ TeV with the data sample collected in Run 2 and corresponding to the integrated luminosity of $L = 2~\textup{nb}^{-1}$. According to estimates performed by M. Dynda\l~et al. under assumption of 5\% systematic uncertainties, ATLAS and CMS will be able to set the following limits on $a_\tau$: $-0.021 \lesssim a_\tau \lesssim 0.017$ (95\%~CL). SMEFT-based projections by Beresford and Liu appear to be more optimistic. In spite of these differences in the projections, one can expect at least a factor of 2 improvement of DELPHI results. Higher statistics foreseen in LHC Run 3 and 4 and reduced systematic uncertainties will allow ATLAS and CMS to push the limits further.

Typical tau decay products have low transverse momenta, therefore most of the ditau events are inaccessible for reconstruction in ATLAS and CMS at least with existing trigger thresholds. The low transverse momentum region can be accessed by the ALICE and LHCb experiments discussed in the following sections.

\section{ALICE potential}

The ALICE experiment provides an opportunity to extend the measurement of ditau events down to low transverse momenta of decay leptons in the pseudorapidity range $|\eta|<0.9$~\cite{ALICE:2008ngc}. During the LHC Run 3 and Run 4, ALICE is going to collect data in the continuous readout mode and accumulate an integrated luminosity of about 2.7 nb$^{-1}$ per one month of Pb--Pb data taking~\cite{Citron:2018lsq}. Energy loss measurements in the ALICE Time Projection Chamber (TPC) can be used for charged particle identification and selection of leading electrons with $p_{\rm T}>0.3$~GeV/$c$ from one tau decay accompanied by one pion or muon from another tau decay. The triggerless data taking will allow ALICE to explore the kinematic region, which is difficult to access in ATLAS and CMS.

In order to study the sensitivity of leading electron $p_{\rm T}$-differential yields to the value of the $a_\tau$ value, we performed Monte Carlo simulations of the 	$\rm {Pb+Pb}\to \rm{Pb+Pb}+\tau\tau$ process at $\sqrt{s_{\rm NN}} = 5.02$ TeV with a dedicated Upcgen generator~\cite{Burmasov:2021phy} that allows one to simulate dilepton pair production in UPCs with an arbitrary value of the magnetic moment. The Upcgen generator is based on the generalized vertex approach used by M. Dynda\l~et al.~\cite{DYNDAL2020135682}. In this approach, the amplitude of the $\gamma\gamma\to\ell\ell$ process is expressed as:
\begin{eqnarray}
	&{\mathcal M}=(-i)\,
	\epsilon_{1 \mu}
	\epsilon_{2 \nu}
	\,\bar{u}(p_{3}) 
	\Big(
	i\Gamma^{(\gamma \ell\ell)\,\mu}(p_1)
	\frac{i(\cancel{p}_{t} + m_{\ell})}{p_t^2 - m_{\ell}^2+i\epsilon}
	i\Gamma^{(\gamma \ell\ell)\,\nu}(p_2) \nonumber\\
	&
	+
	i\Gamma^{(\gamma \ell\ell)\,\nu}(p_2)
	\frac{i(\cancel{p}_{u} + m_{\ell})}{p_u^2 - m_{\ell}^2+i\epsilon}
	i\Gamma^{(\gamma \ell\ell)\,\mu}(p_1) \Big)
	v(p_{4}) \,.
	\label{eq:amplitude}
\end{eqnarray}
Here, $p_1$ and $p_2$ are the four-momenta of the photons, $p_3$ and $p_4$ are the four-momenta of the produced leptons, $\epsilon_{1 \mu}$ and $\epsilon_{2 \nu}$ are the polarization vectors of the photons, $p_t = p_2 - p_4$, $p_u = p_1 - p_4$, $\Gamma^{(\gamma \ell\ell)}$ is the generalized vertex function~\cite{DYNDAL2020135682}, which depends on the magnitude of the momentum transfer $q$:
\begin{equation}
	i\Gamma^{(\gamma \ell\ell)}_{\mu}(q) = 
	-ie\left[ \gamma_{\mu} F_{1}(q^{2})+ \frac{i}{2 m_{\ell}}
	\sigma_{\mu \nu} q^{\nu} F_{2}(q^{2})
	\right]
	\,,
	\label{eq:vertex}
\end{equation}
where $m_\ell$ is the lepton mass, $F_1(q^2)$ and $F_2(q^2)$ are the Dirac and Pauli form factors equal to $F_1(0)=1$ and $F_2(0)=a_\ell$ in the asymptotic limit $q^2 \to 0$. Photons emitted by relativistic heavy ions are close to this limit since the photon virtuality is of the order of $(\hbar c/R_A)^2 \sim 10^{-3}$~GeV$^2$ where $R_A$ is the nuclear radius. 

The cross section of the dilepton production process $\rm {Pb+Pb}\to \rm{Pb+Pb}+\ell\ell$ is then obtained by convolution of the elementary $\gamma\gamma\hspace{-1mm}\to\hspace{-1mm}\ell\ell$ cross section and the two-photon luminosity $\mathrm{d} N^2_{\gamma\gamma}/\mathrm{d} Y \mathrm{d} M$:
\begin{equation}
	\frac{\mathrm{d^2}\sigma(\mathrm{Pb\hspace{-1mm}+\hspace{-1mm}Pb}\hspace{-1mm}\to\hspace{-1mm}\mathrm{Pb\hspace{-1mm}+\hspace{-1mm}Pb}\hspace{-1mm}+\hspace{-1mm}\ell\ell)}{\mathrm{d} Y \mathrm{d} M} = \frac{\mathrm{d}^2 N_{\gamma\gamma}}{\mathrm{d} Y \mathrm{d} M} \sigma(\gamma\gamma\hspace{-1mm}\to\hspace{-1mm}\ell\ell)\,,
	\label{eq:nuc_cs}
\end{equation}
where $Y$ and $M$ are the rapidity and invariant mass of the dilepton pair. Tau decays into final state products were performed with the Pythia8 decayer~\cite{pythia8}.

\begin{figure}[h]
	\centering
	\includegraphics[width=0.65\textwidth]{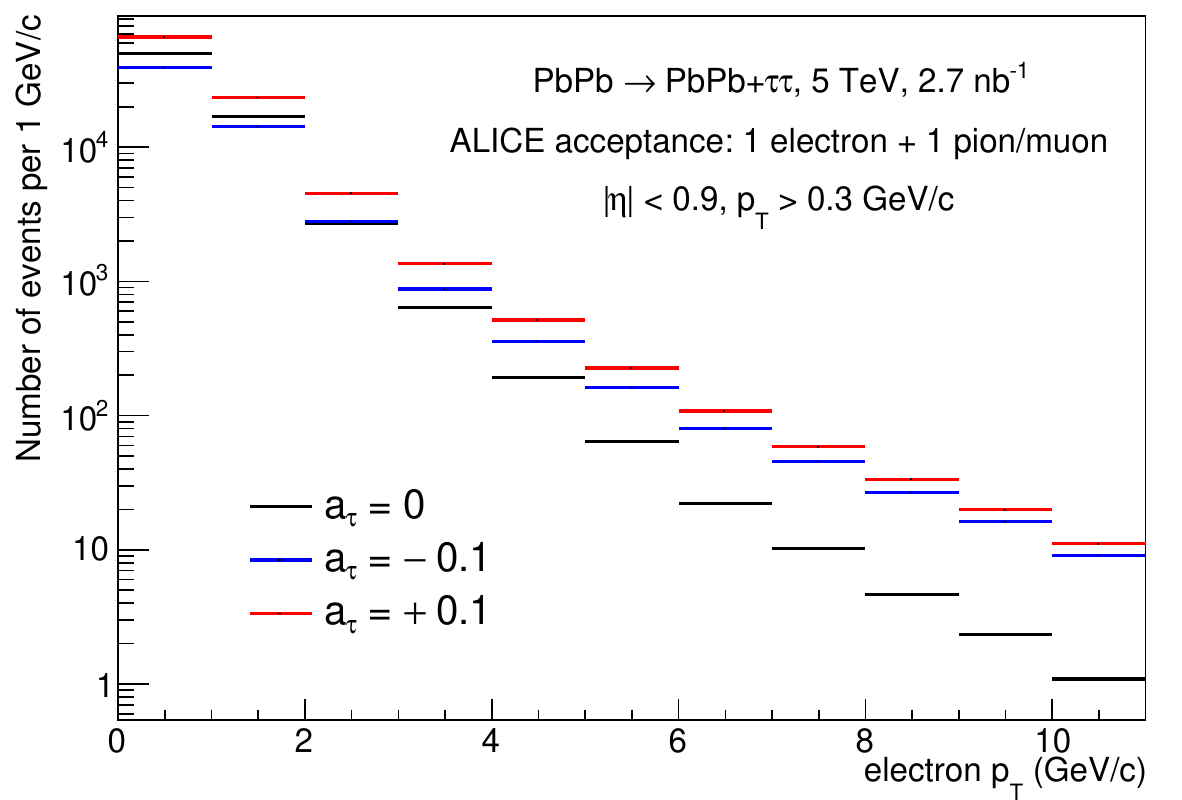}
	\caption{Leading electron $p_{\rm T}$  distributions for three different values of $a_\tau$ in ${\rm Pb+Pb}\to{\rm Pb+Pb +\tau\tau}$ events at $\sqrt{s_{\mathrm{NN}}}=5.02$~TeV corresponding to the integrated luminosity of $L = 2.7~\textup{nb}^{-1}$.}
	\label{fig:electronPtDistr}
\end{figure}

Fig.~\ref{fig:electronPtDistr} illustrates an example of leading electron $p_{\rm T}$  distributions for three different values of $a_\tau$ in ${\rm Pb+Pb}\to{\rm Pb+Pb +\tau\tau}$ events at $\sqrt{s_{\mathrm{NN}}}=5.02$~TeV corresponding to the integrated luminosity of $L = 2.7~\textup{nb}^{-1}$ that is expected in the first year of the LHC Run 3. According to this figure, the ALICE experiment will be able to accumulate about 70000 ditau event candidates, an order of magnitude higher than the number of ditau events collected by ATLAS and CMS in the LHC Run 2~\cite{Beresford,DYNDAL2020135682}.

One can also notice that the leading electron $p_{\rm T}$ dependence on $a_\tau$ is different at low and high transverse momenta. At electron $p_{\rm T}$ above 3 GeV/$c$, the distributions for $a_\tau=0.1$ and $a_\tau=-0.1$ lie above the distribution corresponding to $a_\tau=0$, while at lower transverse momenta the spectrum for $a_\tau=-0.1$ is below the $a_\tau=0$ reference. This nontrivial behaviour of the electron spectra for different $a_\tau$ values allows to increase the sensitivity to $a_\tau$ via $p_{\rm T}$-differential measurements.

\begin{figure}[h]
	\centering
	\includegraphics[width=0.65\textwidth]{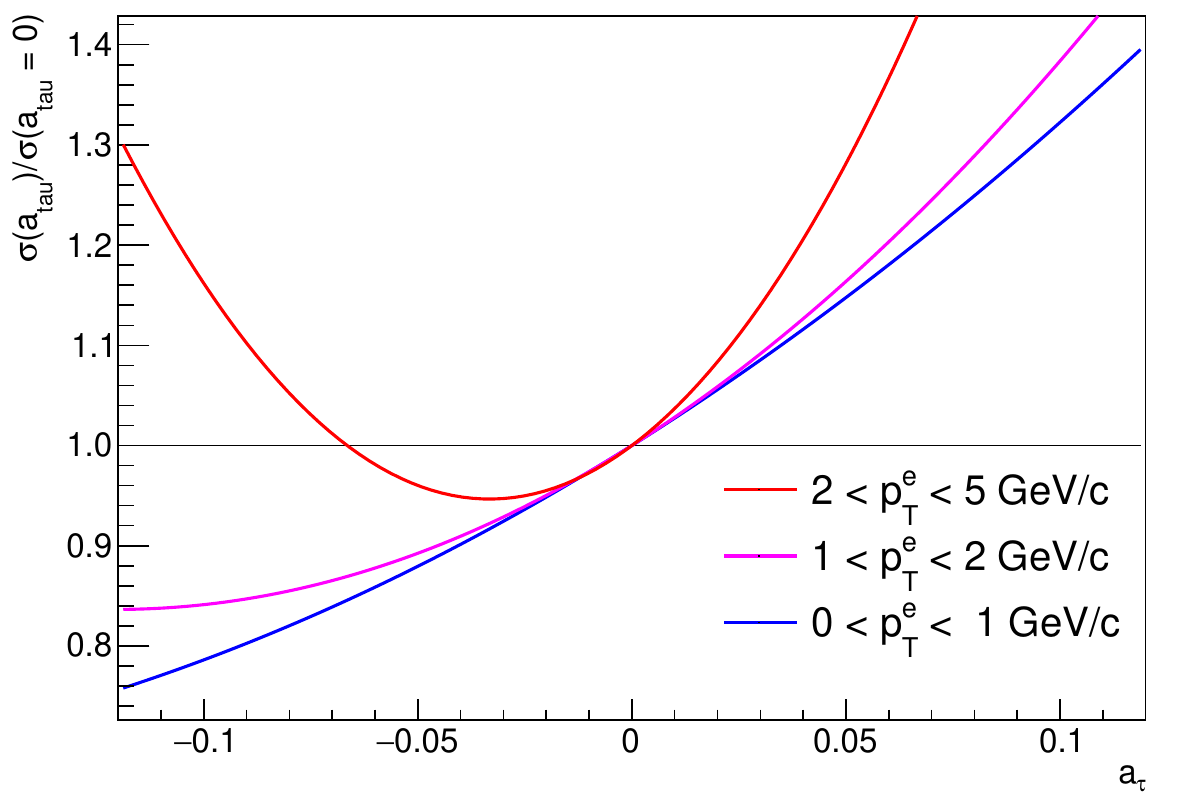}
    \caption{The ratios of electron $p_{\rm T}$ yields as a function of $a_\tau$ relative to the yields at $a_\tau=0$ in several $p_{\rm T}$ intervals.}
	\label{fig:cs_vs_atau}
\end{figure}

The sensitivity to $a_\tau$ can be studied in more detail using the ratios of electron $p_{\rm T}$ yields  as a function of $a_\tau$ relative to the $a_\tau=0$ reference in several $p_{\rm T}$ intervals. These ratios are shown in Fig.~\ref{fig:cs_vs_atau} for three $p_{\rm T}$ ranges in which the statistical uncertainty of the yield is expected to be about 1\% for the data sample that will be collected by ALICE in the first year of Run~3. These ratios reveal a parabolic dependence on $a_\tau$ in the vicinity of $a_\tau=0$ with higher $p_{\rm T}$ ratios showing a steeper raise. In particular, higher $p_{\rm T}$ intervals appear to be more sensitive to positive $a_\tau$ values while low $p_{\rm T}$ intervals ensure better sensitivity for negative $a_\tau$.

\begin{figure}[t]
	\centering
	\includegraphics[width=0.7\textwidth]{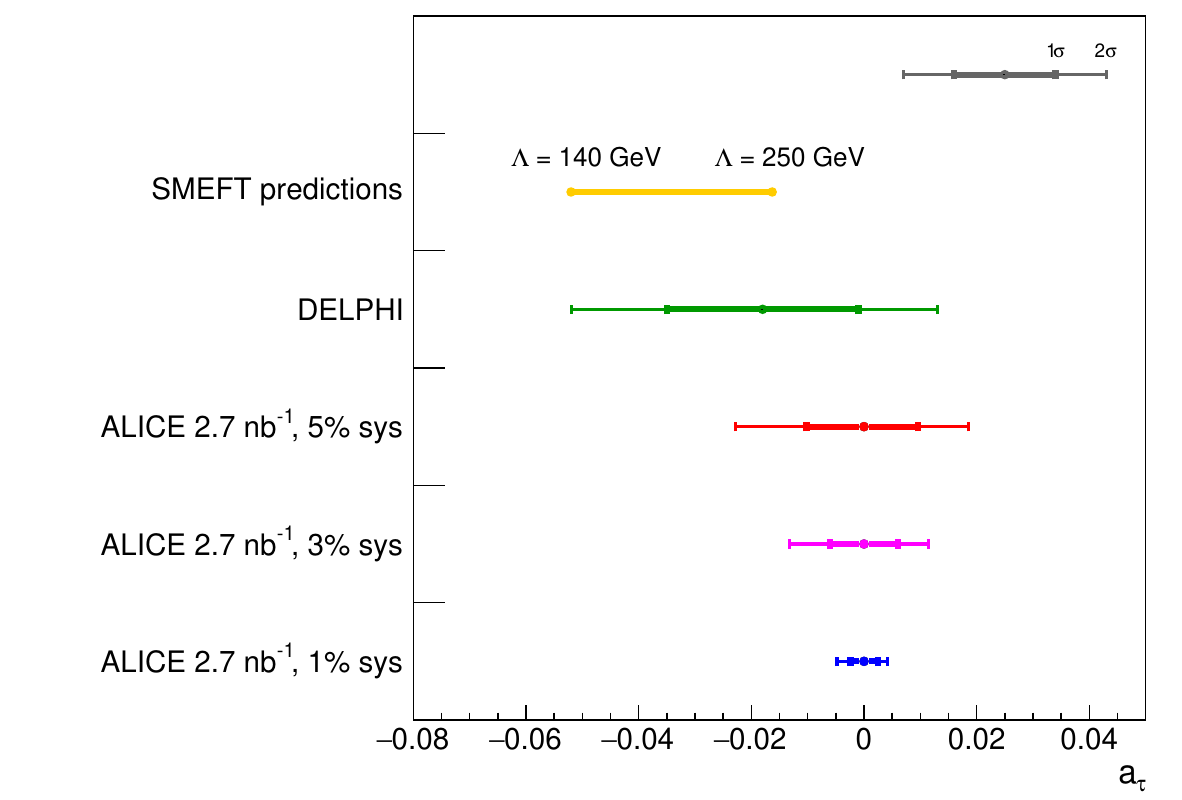}
	\caption{Expected 68\% and 95\% CL limits on $a_\tau$ measurements with the ALICE experiment for three different assumptions on systematic uncertainty (1\%,~3\%,~5\%) in comparison with DELPHI results~\cite{DELPHITauLimits} and SMEFT predictions for a composite $\tau$ scenario from~\cite{Beresford}.}
	\label{fig:aliceLimits}
\end{figure}

Possible limits on $a_\tau$, which can be set by the ALICE collaboration with the data sample from the first year of Run 3, were estimated using the following $\chi^2$ definition:
\begin{equation}
    \chi^2(a_\tau) = \sum_{i,\rm bins} \frac{[S_i(a_\tau) - S_i(0)]^2}{S_i(0)+\zeta S_i^2(0)}
\end{equation}
where $S_i(a_\tau)$ is the yield of electrons in the $p_{\rm T}$ range $i$, $\zeta$ is the expected level of the relative systematic uncertainty. Expected limits on $a_\tau$ at 68\% and 95\% CL are shown in Fig.~\ref{fig:aliceLimits} for three different assumptions on the systematic uncertainty $\zeta$ (1\%,~3\%,~5\%) in comparison with DELPHI results and SMEFT predictions corresponding to the range of confinement scales in a composite $\tau$ scenario~\cite{Beresford}.

The obtained projections indicate that the ALICE experiment will be able to improve \mbox{DELPHI} limits by at least a factor of 2 for the most pessimistic scenario of 5\% systematic uncertainties. Moreover, the level of systematic uncertainties appears to be decisive for the ALICE case. In order to reduce the impact of systematic uncertainties, one can consider the measurement of the $p_{\rm T}$--differential electron yields relative to the dilepton production cross section $\gamma\gamma \to \ell\ell$ in the same UPC data sample as was proposed in~\cite{DYNDAL2020135682}.

\section{LHCb potential}

The LHCb detector is a single-arm spectrometer covering the forward pseudorapidity range $2 < \eta < 4.5$ and allowing for isolation of electrons from charged hadrons at momenta above 2 GeV/$c$ and reconstruction of muons with momenta above 6 GeV/$c$~\cite{LHCb:2014set}. Though the LHCb experiment is not designed to cope with high occupancy of central PbPb collisions, it was able to record UPC events using a dedicated trigger on muons with $p_{\rm T} > 0.9$ GeV/$c$~\cite{LHCb:2021bfl}.

\begin{figure}[htb]
	\centering
	\includegraphics[width=0.65\textwidth]{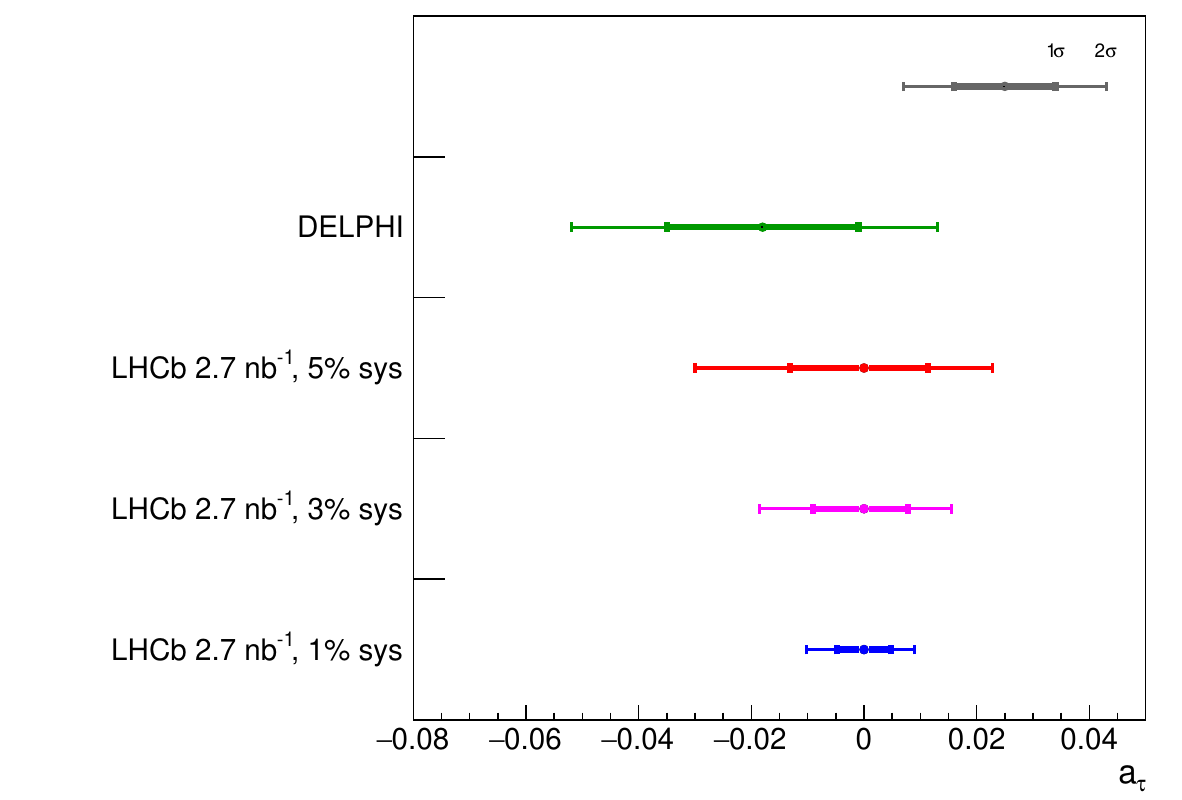}
	\caption{Expected 68\% and 95\% CL limits on $a_\tau$ measurements with the LHCb experiment for three different assumptions on systematic uncertainty (1\%,~3\%,~5\%) in comparison with DELPHI results~\cite{DELPHITauLimits}.}
	\label{fig:lhcbLimits}
\end{figure}

We repeated the analysis described in the previous section for the LHCb case considering a possibility to reconstruct ditau event candidates with one muon ($p > 6$ GeV/$c$) accompanied by one electron ($p > 2$ GeV/$c$) in the pseudorapidity range $2 < \eta < 4.5$ and assuming minimum muon $p_{\rm T}$ threshold of 0.9 GeV/$c$. These kinematic constraints and typical integrated luminosity of 2 nb$^{-1}$ would allow LHCb to record about 7000 ditau event candidates. The expected limits on $a_\tau$ for this statistics are shown in Fig.~\ref{fig:lhcbLimits} for different assumptions on the total systematic uncertainty.

Though possible constraints on $a_\tau$ are expected to be weaker compared to other LHC experiments, the LHCb experiment will be able to perform the analysis in the clean electron-muon channel and in the complimentary rapidity region thus providing a possibility to reduce uncertainties of a cross-experimental measurement.

\section{Conclusion}

In conclusion, ultra-peripheral collisions at the LHC provide a promising tool to measure the anomalous magnetic moment of the $\tau$ lepton. We reviewed recent proposals for the $a_\tau$ measurement in ATLAS and CMS with the available $\rm {Pb+Pb} \to \rm {Pb+Pb}+\tau\tau$ event sample and provided our projections for ALICE and LHCb experiements.

The obtained results indicate that the currently recorded dataset in ATLAS and CMS and future ALICE and LHCb data can surpass DELPHI limits by at least a factor of two. The ALICE experiment has a potential to extend the transverse momentum reach of leading electrons from $\tau$ decays down to low $p_{\rm T}$ of about 0.3 GeV/$c$ while the LHCb experiment may provide a measurement in the complimentary rapidity region not accessible by other experiments.

The $a_\tau$ measurement via ditau production in UPCs requires careful evaluation of various theoretical uncertainties related to higher order effects~\cite{HeckenRad,SpencerRad} and precision of the equivalent photon approximation used to calculate the effective two-photon luminosity in ultra-peripheral collisions. These uncertainties will be addressed in the future studies.

\section*{Acknowledgements}
This work was supported by the Russian Foundation for Basic Research according to the project no.~21-52-14006 and the Austrian Science Fund according to the project no.~I~5277-N.

\bibliographystyle{SciPost_bibstyle}
\bibliography{bibliography.bib}

\nolinenumbers

\end{document}